\documentstyle{ioplcqg}

\makeatletter

\expandafter\ifx\csname documentclass\endcsname\relax
  \def\@jscquest{{\bf ?}}
  \long\def\@firstofone#1{#1}
  \let\G@refundefinedtrue\relax
  \let\@latex@warning\@warning
\else
  \def\@jscquest{\leavevmode\mbox{\reset@font\bfseries ?}}
\fi
 
\let\@internalcite\cite
\def\@normalcitesep{;\penalty-\@m\ }%
\def\@ccitesep{, }%
\def\cite{\let\@citesep\@normalcitesep
 \def\@cite##1##2{(\nobreak\hskip 0in{##1\if@tempswa , ##2\fi})}%
 \def\citeauthoryear##1##2{##1,\penalty\@m\ ##2}\@internalcite}
\def\ccite{\let\@citesep\@ccitesep
 \def\@cite##1##2{{##1\if@tempswa , ##2\fi}}%
 \def\citeauthoryear##1##2{##1\penalty\@m\ (##2)}\@internalcite}
 
\def\citeauthor#1{%
 \def\citeauthoryear##1##2{##1}\@citedata{#1}}
\def\citeyear#1{%
 \def\citeauthoryear##1##2{##2}\@citedata{#1}}
 
\def\@citedata#1{%
 \if@filesw\immediate\write\@auxout{\string\citation{#1}}\fi
 \@ifundefined{b@#1}{\@jscquest	
   \G@refundefinedtrue\@latex@warning
   {Citation `#1' on page \thepage \space undefined}}%
  {\csname b@#1\endcsname}}
 
\def\@citex[#1]#2{%
  \let\@citea\@empty
  \@cite{\@for\@citeb:=#2\do
    {\@citea\let\@citea\@citesep
     \edef\@citeb{\expandafter\@firstofone\@citeb}%
     \if@filesw\immediate\write\@auxout{\string\citation{\@citeb}}\fi
     \@ifundefined{b@\@citeb}{\@jscquest	
       \G@refundefinedtrue\@latex@warning
       {Citation `\@citeb' on page \thepage \space undefined}}%
     {\csname b@\@citeb\endcsname}}}{#1}}

\def\@biblabel#1{}
\def\thebibliography#1{\section*{References}
  \addcontentsline{toc}{section}{References}
  \list{}{\labelwidth\z@
    \leftmargin 1.5pc
    \itemindent-\leftmargin}
    \footnotesize
    \parindent\z@
    \parskip\z@ plus .1pt\relax
    \def\newblock{\hskip .11em plus .33em minus .07em}
    \sloppy\clubpenalty4000\widowpenalty4000
    \sfcode`\.=1000\relax}

\makeatother

\font\SYM=msbm10 
\def\Real{\hbox{\SYM R}}
\newcommand{\smfrac}[2]{{\textstyle{#1\over#2}}}
\def\half{\smfrac{1}{2}}
\def\arctanh{\mathop{\rm arctanh}\nolimits}
\def\arctan{\mathop{\rm arctan}\nolimits}

\begin{document}

\title{Stationary and static cylindrically symmetric
Einstein spaces of the Lewis form}%
[Stationary cylindrical Einstein spaces]

\author{
M A H MacCallum\dag\ddag\protect\ftnote{4}%
{E-mail address: M.A.H.MacCallum@qmw.ac.uk} and
N O Santos\dag\S\protect\ftnote{5}{E-mail address: nos@on.br}}%
[M.A.H. MacCallum and N.O.Santos]

\address{\dag\ School of Mathematical Sciences,
Queen Mary and Westfield College,
Mile End Road, London E1 4NS, UK}

\address{\ddag\ Department of Mathematics and Applied Mathematics,
University of Cape Town, Rondebosch 7700, South Africa}

\address{\S\ Departamento de Astrof\'{\ii}sica,
CNPq-Observat\'orio Nacional,
Rua General Jos\' e Cristino 77,
20921-400, Rio de Janeiro, RJ, Brazil}

\recd{\today}

\begin{abstract}
The derivation of the general solutions for stationary and static
cylindrically symmetric Einstein spaces of Lewis form is revisited and
the physical and geometrical meaning of the parameters appearing in
the resulting solutions are investigated. It is shown that three of
the parameters (and the value of the cosmological constant) are essential, of
which one characterizes the local gravitational field and appears in
the Cartan scalars, while the remaining two give information about the
topological identification made to produce cylindrical symmetry. Other
than the cosmological constant, they can be related to the parameters
of the vacuum Weyl and Lewis classes, whose interpretation was
previously investigated by da Silva et al.\ (1995a, 1995b).
\pacs{0420J, 0440N}
\end{abstract}

\eqnobysec

\section{Introduction} \setcounter{equation}{0}

In a recent paper, \ccite{San93} gave a set of solutions for
stationary cylindrically symmetric spacetimes of the Lewis form
(\citeyear{Lew32}) with a cosmological constant. Krasinski had
previously given these in a different form (\citeauthor{Kra75}
\citeyear{Kra75}, \citeyear{Kra94}). Both use coordinates in which the
static limit is hard to obtain. We shall reconsider the derivation of
the complete set of stationary and static solutions. In another set of
recent papers, da Silva et al.\ (\citeyear{SilHerPai95},
\citeyear{SilHerPai95a}) have given some physical identifications for
the parameters of the corresponding vacuum solutions. Here we extend
that investigation to the cases with cosmological constant.

We first make some remarks on the problem of definition of cylindrical
symmetry. In \ccite{KraSteMac80} a spacetime is defined to be
cylindrically symmetric if it is axisymmetric about an infinite axis
and translationally symmetric along that axis.  The definition of
axisymmetry gives some difficulties (\citeauthor{MarSen93}
\citeyear{MarSen93}, \citeyear{MarSen95}). In the non-singular case,
we can characterize axisymmetry by the requirement that some Killing
vector (a) has closed spacelike orbits near the axis and (b) has
length zero at the axis. In order for the axis to be non-singular we
require elementary flatness at the axis.  If such an axisymmetric
spacetime is also stationary, then one can prove that the stationary
and axisymmetry Killing vectors commute and form surfaces with
orthogonal transitivity, but these results depend on the existence of
regular axis points (see chapter 17 of \ccite{KraSteMac80}).

These ideas are not currently well-defined if the axis is not regular,
or has no regular part, but to exclude metrics without regular axes
would be too restrictive as it excludes many of the known solutions
which one would want to call cylindrical. In particular it rules out
solutions which if continued to the axis ($\rho=0$ in the usual
coordinate systems) would have strange behaviour but which can be
joined at some $\rho = \rho_0 > 0$ to a perfectly regular and well-behaved
interior solution. This
requirement, together with the existence of two commuting
spacelike Killing vectors of which one has closed orbits and the other
has non-closed orbits, so that they are transitive on surfaces of
topology $S^1 \times \Real^1$, could be taken as the definition of
cylindrical symmetry. However, at present we
know no way to test whether a metric
could form a part of a globally well-behaved cylindrically symmetric
solution in the sense of \citeauthor{KraSteMac80} and it is therefore simpler
to adopt the less demanding definition requiring only the existence of
spacelike surfaces of symmetry with topology $S^1 \times \Real^1$.

If there is no regular axis, one could define the presence of a
singular axis by requiring the spacelike Killing vector with closed
orbits to have a length which tends to zero, but the subsequent
treatment of such spacetimes has some difficulties. In particular one
cannot prove the orthogonal transitivity property, either in the
stationary case (where it can occur with a timelike surface of
transitivity), or in the more general non-stationary cylindrically
symmetric case, in which \ccite{KraSteMac80} assume it to occur with a
spacelike surface of transitivity (see equation (20.1) there). In
particular, if in the latter form one assumes that the metric
coefficients are independent of time, one would wish to call the
solution static even though the metric does not take the usual form
for a stationary axisymmetric metric with orthogonal transitivity: for
some such solutions see \ccite{ChiGuvNut75,Mac83}. Another possible
problem is shown by the solution discussed by \ccite{Lem94} in which the
``axis'' $\rho =0$ in our coordinates is interpreted as the horizon of
a cylindrical black hole.

Thus the Lewis metric form used by Santos does not include all
possible stationary Einstein metrics with cylindrical symmetry,
defined as above. Even in the vacuum case the solutions of the Lewis form
in general do not have well-defined axes (where the Killing vector
with closed orbits has zero length), or if they do the axis is not in
general elementarily flat.  We therefore begin by simply assuming the
Lewis form of the metric \cite{Lew32},
\begin{equation}
\label{Lewisform}
\d s^2=-f \d t^2 +2k\,\d t\,\d\varphi +\ell\,\d\varphi^2
+ \e^\nu \d r^2 + \e^\mu \d z^2
\end{equation}
where $f$, $k$, $\ell$, $\nu$ and $\mu$ are functions of
$r$ only, and the ranges of the coordinates $t$, $r$, $z$ and $\varphi$
are initially taken to be
\begin{equation}
\fl
-\infty<t<\infty \qquad -\infty<r<\infty \qquad -\infty<z<\infty\qquad
0\leq \varphi \leq 2\pi
\label{Coords}
\end{equation}
with the hypersurfaces $\varphi=0$ and $\varphi=2\pi$ identified to
ensure the cylindrical symmetry. Note that we have not assumed $0 \leq
r$, in order both to avoid assuming the existence of an axis and to
allow more freedom of coordinate choice. The coordinates are numbered
\begin{equation}
x^0=t\tqs x^1=r\tqs x^2=z \tqs x^3=\varphi\,.\label{2.3}
\end{equation}

We should note immediately that the identification made on $\varphi$
makes this an improper coordinate system: it does not obey the
differential geometric requirement of giving a homeomorphism between a
region of the manifold and an open set in $\Real^4$.  For that reason,
we should really use at least two coordinate patches to cover the
whole manifold. The permissible local transformations of coordinates,
on these two or more patches, which
are compatible with a identification in
$\varphi$, with period $P^\prime$ say, are \cite{Mac97}
\begin{equation}
\bar{t} = T_1 t + T_0, \quad \bar{r} = \bar{r}(r), \quad \bar{z} = Z_1
z + Z_0, \quad \bar{\varphi} = X_1 t + X_2 \varphi + X_0,
\end{equation}
where the upper case letters denote constants, and $\bar{r}$ is an
arbitrary function. It is clear that the
changes of origin of $t$, $z$ and $\varphi$ do not alter any metric
coefficient and can be ignored, so we need only consider the more
restricted transformations
\begin{equation}
\label{transf1} 
\bar{t} = T_1 t ,  \quad \bar{r} = \bar{r}(r), \quad \bar{z} = Z z ,
\quad \bar{\varphi} = X_1 t + X_2 \varphi .
\end{equation}
However, we could also locally allow the more general form
\begin{equation}
\label{transf2}
\bar{t} = T_1 t + T_2 \varphi,  \quad \bar{r} = \bar{r}(r), \quad
\bar{z} = Z z , \quad \bar{\varphi} = X_1 t + X_2 \varphi .
\end{equation}
In \ccite{Mac97}, it is shown that of the four parameters $T_1$,
$T_2$, $X_1$ and $X_2$ only two are essential; they fix the
topological identification needed to reach the actual solution
starting from its locally equivalent standard form. If in the above
transformations the barred coordinates refer to this standard form then
the essential parameters $S$ and $P$ of \ccite{Mac97} are
$-X_2/T_2$ and $2\pi X_2$ (or we could say, more succinctly, that
$X_2$ and $T_2$ are essential).

The radial coordinate can be chosen so that $\nu=\mu$ (as
\ccite{San93} does initially) or so that $\nu=0$, which is the more
convenient choice of radial coordinate introduced later by
\ccite{San93}, called $r^*$ in that paper, which is in fact the
natural $r$ coordinate arising from the existence of
hypersurface-homogeneity of Bianchi type I on timelike surfaces (which
thus necessarily have geodesic normal unit vectors). With this choice the
remaining coordinate freedom of $r$ is $\bar{r} = \pm r+R$ where $R$ is a
constant.

\section{The Einstein space solutions}

We now revisit and complete the derivation of the solutions of the
Einstein equations
\begin{equation}
R_{ab} = \Lambda g_{ab}
\end{equation}
for the metric form considered.  Following \ccite{San93} we find that
if $u=f/\ell$ and $v=k/\ell$ then with the choice of $r$ given by
$\nu=\mu$, and denoting $\d /\d r$ by ${}^\prime$, we have
\begin{equation}
u^\prime = \Theta^\prime (u+v^2), \qquad v^\prime = \Phi^\prime
(u+v^2)
\end{equation}
where $(\rho \Theta^\prime)^\prime= 0 = (\rho \Phi^\prime)^\prime$,
$\rho$ (denoted $D$ in \ccite{San93}) being the usual radius parameter
$\rho^2 = f\ell+k^2$ (which in vacuum and some other cases can also be
taken as an isothermal coordinate in the $(r,\,z)$ surfaces). There
are now several possibilities. If both $\Theta^\prime$ and
$\Phi^\prime$ are non-zero, they must be multiples of one another, say
$\Phi^\prime=A\Theta^\prime$, and this leads (as in \ccite{San93}) to
$v=Au+B$ where $A$ and $B$ are constants, i.e.\ to $k=Af+B\ell$. If
one of $\Theta^\prime$ and $\Phi^\prime$ is zero, so that either
$\Theta$ or $\Phi$ is constant (possibly zero), then again $f$, $k$
and $\ell$ are linearly dependent, and this is also true if $\ell=0$
so that $u$ and $v$ are not well-defined. Thus in all cases the
proposition of \ccite{Mac97} applies, and {\em all solutions are
locally equivalent either to a diagonal metric, or one with a null
Killing vector, or a `windmill' form (which can be considered as a
complex diagonal metric)}. Moreover, the solution for the diagonal
case will give the solutions for the remaining cases if we allow
complex values for parameters and coordinates and take limits as in
\ccite{McI92}.

Taking the form (\ref{Lewisform}) with $k=0=\nu$ we find (cf.\
\ccite{Lin85} and \ccite{San93}) that if $G = \rho e^{\mu/2}$ then
\[ G^{\prime\prime} + 3\Lambda G = 0. \] The distinct solutions (after
choice of the origin and sign of $r$) are
\begin{eqnarray}
\label{minus1}
G  &=C\cosh\left( \sqrt {3\left\vert {\Lambda} \right\vert } r \right)
 ,& \quad \Lambda <0;\\
\label{minus2}
 &=C \exp \left( \sqrt {3\left\vert {\Lambda } \right\vert }r \right)
 ,& \quad \Lambda <0;\\
\label{minus3}
 &=C\sinh\left( \sqrt {3\left\vert {\Lambda } \right\vert }r \right)
 ,& \quad \Lambda <0;\\
\label{zero}
G &=C r,& \quad \Lambda =0;\\
\label{plus}
G &=C\sin\left( \sqrt {3\Lambda }r \right) ,&\quad \Lambda >0,
\end{eqnarray}
where in all cases $C > 0$. The interpretation of $G < 0$ is
unclear. For example,  it might represent a double-covering of the
spacetime or a region interior to a horizon.
The vacuum solutions are included for comparison purposes although these
solutions, which can all be related to the Kasner form (see
\ccite{McI92} and references therein), have been discussed in many
earlier papers.  By scaling the three ignorable
coordinates we can arrive at $C = 1/\sqrt{\vert 3\Lambda \vert}$ if
$\Lambda \neq 0$: for $\Lambda = 0$ we can normalize so that $C=1$, and we
denote these normalized cases by $G_0$. They have the property that
\[ (G_0^\prime)^2 + 3\Lambda (G_0)^2 = \eta \]
where in the five cases (\ref{minus1}--\ref{plus}) $\eta$ is
respectively $-1$, $0$, $1$, $1$ and $1$.

The case $\Lambda < 0$ gives three forms of solution, but
\ccite{Lin85} considers only (\ref{minus3}), and the differences
are not obvious in Krasinski's complicated form. In \ccite{San93} all
three are included in the general form
\begin{equation}
G =C_1\cosh\left( \sqrt {-3\Lambda} r \right)+
C_2\sinh\left( \sqrt {-3\Lambda} r
\right) ,
\end{equation}
the three cases corresponding to $\vert C_1\vert > \vert C_2\vert $,
$\vert C_1\vert = \vert C_2\vert $, and $\vert C_1\vert < \vert
C_2\vert $. In the first and third cases the $C$ above is
$\sqrt{\vert C_1^2-C_2^2 \vert}$, while in the second case $C=\vert
C_1 \vert$. The origin shifts in $r$ required to reach (\ref{minus1})
and (\ref{minus3}) are respectively $\arctanh(C_2/C_1)$ and
$\arctanh(C_1/C_2)$.
In the
similar formulae for the case
(\ref{zero}) $C=C_2$ and the shift is $C_1/C_2$ and in the case (\ref{plus})
$C=\sqrt{C_1^2+C_2^2}$ and $R=\arctan(C_1/C_2)$.

With these specializations, and using
allowable scalings of the remaining coordinates, we can write the
resulting metric \cite{Lin85} as
\begin{equation}
\label{diag}
\d s^2= \d r^2 + G_0^{2/3}[\sum_{j=0,2,3} \varepsilon_j \exp (2(p_j-1/3)
U) (\d x^j)^2]
\end{equation}
where $\varepsilon_j$ is $1$ except for $x^0=t$, for which it is $-1$,
and $\d U/\d r = 1/G_0$. (No constant of integration is given in $U$
as it can be absorbed by rescalings of coordinates, but the rescalings
allowed appear explicitly in the discussion below.) The constants
$p_j$ (which are $1/3+KK_j/2$ in terms of Linet's notation) obey
\begin{equation}
\label{Kascond}
 \sum_j p_j = 1, \qquad \sum_j p_j^2 = (2\eta+1)/3.
\end{equation}

The parametrization here has been chosen to agree with the usual forms
of the Kasner vacuum metric (equations (11.50) and (11.51) in
\ccite{KraSteMac80}) and its `windmill' counterpart \cite{McI92} for
the case $\Lambda = 0$, where $G=r = \exp U$. As in the Kasner vacuum
case, a complex coordinate change gives related `cosmological'
solutions, i.e.\ solutions for which the essential variable has a timelike
character. These related metrics have been given several times: Kasner
himself (\citeyear{Kas25}) gave the case related to (\ref{plus}),
\ccite{Sau67} (see equation (11.52) in \ccite{KraSteMac80}) the cases
related to (\ref{minus3}) and (\ref{plus}), \ccite{Kal70} the case
related to (\ref{minus1}), \ccite{Spi79} gave the full set, and there
may be other occurrences we have not found.  The metric forms are
included in those considered by \ccite{Kel75}, but he seems not to
give explicitly the solutions for a cosmological
constant. \ccite{Lin85} only considered
the forms with (\ref{minus3}) and (\ref{plus}), as he wished to have
an axis $\rho = 0$ at some $r$.

Note that for $\eta = 0$ the only real solution is $p_i=1/3$ for all
$i$: other cases with $\eta=0$, and all cases with $\eta=-1$, will
have complex values of $p_i$. It is not obvious whether in general the
resulting metrics, with the coordinates also considered to be complex,
will have a real section, as happens in the `windmill' case
\cite{McI92}. However, there are clearly cases where this does happen:
for example, with $\eta=-1$, if $p_0=1$, $p_2=\bar{p_3}=
\sqrt{2}i/\sqrt{3}$ where we get a real solution by
replacing $z$ and $\varphi$ by $x-iy$ and $x+iy$ respectively.  Linet
ignores these cases, taking his $K^2 >0$. They are implicitly present
in \ccite{San93}, where they correspond to complex values of $\delta$
and $\alpha\gamma/\beta$ in his equation (51), as corrected (see the
erratum to the paper and below). The
similar examples with $p_2=1$ are covered by his equation (57). If we
assume that $p_2$ is real and $p_0=a+ib=\bar{p_3}$, we find this is
possible, with non-zero real $b$, for values of $a$ outside $[0,\,2/3]$ if
$\eta=1$, for $a \neq 1/3$ if $\eta=0$ and for any $a$ if
$\eta=-1$. The cases with a null Killing vector can be obtained by
going to coordinates in which the metrics appear stationary and then
taking a suitable limit.

There are clearly two essential parameters in (\ref{diag}), one of the
$p_i$, say $p_2$, and the cosmological constant (and the discrete
parameter $\eta$ if $\Lambda < 0$). One could argue that $\Lambda$
should not be regarded as a parameter, since in the usual Einstein
theory it would be a universal constant. However, if the term is
regarded as arising as a relic of quantum processes, or if we consider
that in writing down the whole family we have an infinite number of
completely disjoint spacetimes, and hence no physical reason for using
the same $\Lambda$, it may make sense to treat $\Lambda$ as a
parameter. To make the solution cylindrical, $\varphi$ must be
periodic, and the period is specified by another essential global
parameter (see \ccite{Mac97}): note
that scaling $\varphi$ to make the period $2\pi$ alters the
$g_{\varphi\varphi}$ of (\ref{diag}) by a constant factor. Thus in the
form (\ref{diag}) for a cylindrical metric with a periodic $\varphi$
there are three parameters, $\Lambda$, $p_2$ and the period (plus the
discrete parameter $\eta$ if $\Lambda < 0$).

For the distinct possible $G_0$ given by (\ref{minus1}--\ref{plus})
the formulae for $U$ are respectively
\begin{eqnarray}
\label{Uminus1}
U &=\arctan \left\vert \sinh \left( \sqrt {3\left\vert {\Lambda }
 \right\vert }r \right) \right\vert
 ,& \quad \Lambda <0;\\
\label{Uminus2}
 &= \exp \left( -\sqrt {3\left\vert {\Lambda } \right\vert }r \right)
 ,& \quad \Lambda <0;\\
\label{Uminus3}
 &=\ln \tanh \half \left( \sqrt {3\left\vert {\Lambda} \right\vert }
 r \right)  ,& \quad \Lambda <0;\\
\label{Uzero}
U &= \ln r,& \quad \Lambda =0;\\
\label{Uplus}
U &=\ln \tan \half \left( \sqrt {3\Lambda }r \right) ,&\quad \Lambda >0,
\end{eqnarray}
cf.\ \ccite{San93}, equations (45)--(49).

We can now obtain the general forms of the corresponding classes of
stationary cylindrically symmetric metrics by applying the
transformations (\ref{transf2}), taking the barred coordinates to be
the ones for the standard form just discussed. This will introduce a
fourth essential parameter (not counting $\eta$) which specifies the
lines in the original $(\bar{\varphi},\,\bar{t})$ surface parallel to
$t=0$ along which the topological identification is to be made, as
well as some inessential parameters: as noted above $X_2$ and $T_2$
can be considered the essential parameters in addition to $\Lambda$
and $p_2$. We do the calculations explicitly only for the diagonal
form, assuming $p_2$ to be real, which leaves us the possibility of
complex $p_0$ and $p_3$. Then
\begin{eqnarray}
\fl
\d s^2 &= & \d r^2 + G_0^{2/3}\left( Z^2 \exp (2(p_2-1/3) U) \d z^2 \right.
\nonumber \\
\fl && + [X_2^2 \exp (2(p_3-1/3)U) - T_2^2 \exp
 (2(p_0-1/3)U)] \d \varphi^2
\label{gensol} \\
\fl && + 2[X_1X_2 \exp
 (2(p_3-1/3)U) - T_1 T_2 \exp (2(p_0-1/3)U)]\d t \d
 \varphi 
\nonumber \\
\fl && \left.- [T_1^2 \exp (2(p_0-1/3)U)- X_1^2 \exp
 (2(p_3-1/3)U) ]\d t^2 \right) \nonumber
\end{eqnarray}
and the full coordinate freedom includes replacing $r$ in the formulae
for $G_0$ and $U$ by $\pm r + R$ where $R$ is a constant. The geometry
has 4 essential parameters $\Lambda$, $p_2$, $X_2$ and $T_2$. 
$G$ is now given by
\[ G^2= Z^2(T_1X_2 - T_2X_1)^2 G_0^2. \]
The remaining free parameters $T_1$, $X_1$, $R$ and $Z$ are
inessential in specifying the geometry of the region covered, but may
be required in order to match this metric to an interior using the
Lichnerowicz conditions, which are the same as the earlier Darmois
conditions but expressed in `admissible coordinates' continuous across
the boundary (see \ccite{BonVic81} and applications in \ccite{Kra75}
and \ccite{BonSanMac97}). This coordinate choice, and therefore the
inessential coordinate parameters, are not required in the Darmois
form, in which the conditions are just that the first and second
fundamental forms of the surface, considered from the two sides, must
agree, and thus only invariantly-defined parameters can appear. Of
course, on trying to match a given solution to the solution here one
may find it has more essential parameters than can in fact be matched
by the solutions here, or parameters which appear in an incompatible
way: there is no guarantee that a match is possible.

These solutions can now be identified with earlier
forms. \ccite{Kra75} used a ``radial'' coordinate $x^2 = k/f$ in the
above notation: this leads to rather complicated forms, and the static
limit $k = 0$ is awkward to handle.  (This coordinate choice is made
in order to fit the exteriors easily to his non-static interior
solutions.) Indeed \ccite{Kra75a} gives only the case $\Lambda = 0$ in
the explicitly static form, as his equation (7.2), and says ``The
generalization of (7.2) to the case $\Lambda \neq 0$ is unexpectedly
very involved, so I do not present it here'', though in fact the
solutions themselves appear in stationary form.  We can identify his
classes, simply by looking at the coefficients in the linear
dependence of $f$, $k$ and $\ell$ (cf.\ \ccite{Mac97}) as follows. His
class A is locally equivalent to the `windmill' form, Class B is
locally equivalent to a real diagonal form, and Classes C and D are
locally equivalent to the form with a null Killing vector, the
difference being that in Class D the identification is made along this
Killing vector's orbit. The identification of Class B with the
(locally) static cases other than his (9.3) is stated in
\ccite{Kra75}, in the paragraph following Table IX, and the proof of
this (which was known to Krasinski at that time (private
communication, 1997)) is hinted at but not completely described in the
paragraph following his equation (9.2).

The version given originally by \ccite{San93} contains some typographical
errors, corrected in the subsequent erratum.

(i) If $K = -\alpha^2 < 0$, we
have
\begin{eqnarray}
{\ell\over \rho}&=&{1\over \alpha}\sinh\left( {{\alpha\Theta} \over {2\beta}}
  \right),\\
{k\over \rho}&=&-\cosh\left( {{\alpha\Theta} \over {2\beta}} \right)
-{\beta \over \alpha}\sinh\left( {{\alpha\Theta} \over {2\beta}} \right),\\
{f\over \rho}&=&-2\beta \, \cosh\left( {{\alpha\Theta} \over {2\beta}}
\right) -{{\beta ^2+\alpha ^2} \over {\alpha}}\sinh\left(
{{\alpha\Theta} \over {2\beta}} \right).
\end{eqnarray}

(ii) If $K =0$, 
\begin{eqnarray}
{\ell\over \rho}&=&{\Theta\over 2\beta} , \\
{k\over \rho}&=&-1 -{\Theta\over 2},\\
{f\over \rho}&=&-{\beta\over 2}\Theta -2\beta.
\end{eqnarray}

(iii) If $K = \alpha^2 > 0$, 
\begin{eqnarray}
{\ell\over \rho}&=&{1\over \alpha }\sin\left( {{\alpha\Theta} \over {2\beta}}
  \right) \\
{k\over \rho}&=&-\cos\left( {{\alpha\Theta} \over {2\beta}} \right)
-{\beta \over \alpha}\sin\left( {{\alpha\Theta} \over {2\beta}} \right),\\
{f\over \rho}&=&-2\beta \cos\left( {{\alpha\Theta} \over {2\beta}} \right)
-{{\beta ^2-\alpha ^2}\over \alpha}\sin\left( {{\alpha\Theta} \over
{2\beta}} \right), 
\end{eqnarray}

where
\[ \Theta =\gamma \Omega + \zeta, \qquad e^{3\mu /2}=\epsilon G
\exp\left( \delta \Omega \right), \qquad \Omega = \int \d r/G .\]
$\delta $, $\gamma $, $\epsilon$ and $\zeta$ are constants of integration.

The 9 constant parameters, $\Lambda$, $C_1$, $C_2$, $\alpha$,
$\beta$, $\gamma$,
$\delta$, $\epsilon$ and $\zeta$, in these solutions are
related, in each case, by one equation, as follows:

(i) $K< 0$
\begin{eqnarray}
 \label{cond1}
3\left\vert \Lambda \right\vert \left(C_1^2-C_2^2\right) &+{1\over 4}
  \delta^2+{{3\alpha^2\gamma^2} \over {16\beta^2}}=&0, \quad \Lambda <0;\\
-C_2^2 &+{1\over 4}\delta^2+{{3\alpha^2\gamma^2} \over {16\beta^2}}=&0,
  \quad \Lambda =0; \\
\label{cond3}
-3\Lambda \left(C_1^2+C_2^2\right) &+{1\over 4}\delta^2+
  {{3\alpha^2\gamma^2} \over {16\beta^2}}=& 0, \quad \Lambda >0.
\end{eqnarray} 

(ii) $K=0$
\begin{eqnarray}
\label{cond4}
3\left\vert \Lambda \right\vert \left(C_1^2-C_2^2\right) &+{1\over 4}
  \delta^2 =&0, \quad \Lambda <0;\\
C_2 &\pm {1\over 2}\delta =&0, \quad \Lambda =0;\\
 -3\Lambda \left(C_1^2+C_2^2\right) &+{1\over 4} \delta^2=&0, \quad
\Lambda >0.
\end{eqnarray}

(iii) $K>0$
\begin{eqnarray}
\label{cond7}
3\left\vert \Lambda \right\vert \left( C_1^2-C_2^2\right) &+{1\over 4}
  \delta^2-{{3\alpha^2\gamma^2} \over {16\beta^2}}=&0, \quad \Lambda<0; \\
-C_2^2&+{1\over 4}\delta^2-{{3\alpha^2\gamma^2} \over {16\beta^2}}=&0,
 \quad \Lambda =0;\\
\label{cond9}
-3\Lambda \left( C_1^2+C_2^2\right) &+{1\over 4}\delta^2-
 {{3\alpha^2\gamma^2} \over {16\beta^2}}=&0, \quad \Lambda >0.
\end{eqnarray}

We now seek to identify the 9 parameters in the form of \ccite{San93}
with the parameters of the general solution (\ref{gensol}).
We have already identified $C_1$ and $C_2$ with $C$ and $R$ above and
we see that $\Omega = U/3\vert \Lambda \vert C $ if $\Lambda \neq 0$,
or $U/C$ if $\Lambda=0$. For the five possible $G_0$ we have
$Z^2(T_1X_2 - T_2X_1)^2$ equal to, respectively $3\vert \Lambda
\vert (C_1^2-C_2^2)$, $3\vert \Lambda \vert C_1^2$, $3\vert \Lambda
\vert (C_2^2-C_1^2)$, $C_2^2$, and $3\Lambda(C_1^2+C_2^2)$. We find
$\delta= \sqrt{3\vert \Lambda \vert} C(3p_2-1)$ if $\Lambda \neq 0$ or
$\delta = C(3p_2-1)$ if $\Lambda=0$, and $\epsilon = Z^2/(T_1X_2 -
T_2X_1)$. 
In case (i) $\alpha\gamma=6 \vert \Lambda \vert
C\beta(p_3-p_0)$ (or $\alpha\gamma=2C\beta(p_3-p_0)$ if $\Lambda=0$);
in case (iii) $p_0$ and $p_3$ are complex conjugates and
$\alpha\gamma=-6i\vert \Lambda \vert C\beta(p_3-p_0)$.
The relations listed as (\ref{cond1}--\ref{cond3}) and
(\ref{cond7}--\ref{cond9}) are then equivalent
to the second of the equations (\ref{Kascond}), the first of these
being already built into the forms above.

The parameters $\alpha$ and $\beta$ are most easily identified from
the linear dependence of $f$, $k$ and $\ell$ which for case (i) is
\[ f - 2\beta k - \beta^2 \ell = -\alpha^2 \ell \]
and can be compared with 
\[ f
+(\frac{T_1}{T_2}+\frac{X_1}{X_2})k-\frac{T_1 X_1}{T_2X_2}\ell
= 0 \]
giving us that $\beta+\alpha$ and $\beta-\alpha$ are $T_1/T_2$ and
$X_1/X_2$ (the correct pairing depending on the choice which keeps
$t$ timelike).
Finally, use of the transformation
\begin{equation}
\label{transf3}
\bar{t} = t - T_2 \varphi/T_1, \quad \bar{\varphi} = -X_1 t/X_2
+ \varphi ,
\end{equation}
on the metric (\ref{gensol}), and the same in terms of $\alpha$ and
$\beta$ on that of Santos, enables us to identify $\zeta$ in
terms of $\ln (X_2/T_1)$.

\section{Invariant properties} \setcounter{equation}{0}

As discussed above, we have 4 essential parameters in the general
stationary metric, including $\Lambda$ but excluding $\eta$. From the
work of \ccite{Mac97} and the results of
\ccite{SilHerPai95,SilHerPai95a} we can expect that 2 of these ($T_2$
and $X_2$) will appear in the holonomy around the curves on which only
$\varphi$ changes, and the remaining two ($\Lambda$ and $p_2$) in
invariants of the local curvature. Further parameters may appear from
coordinate conditions on a matching to an interior solution.

We can compute the ``Cartan scalars'' which completely characterise
the local geometry according to the procedure for invariant
classification described in \ccite{Kar80,PaiRebMac93} and
\ccite{MacSke94}. For case (i) with the $G_0$ of equation
(\ref{minus3}) and $C=1/\sqrt{3\vert\Lambda\vert}$, this calculation,
which was carried out using the computer algebra package CLASSI built
on SHEEP, gave, as the non-trivial classifying invariants, the
cosmological constant and
\begin{equation}\eqalign {\label{CS1}
\Psi_0 = \Psi_4 
\\
\phantom{\Psi_0 }
=   1/4\alpha\beta^{ -1}\gamma\delta\vert\Lambda\vert
      \sinh^{ -2}(\sqrt{3\vert\Lambda\vert}r) \\
\phantom{\Psi_0 }
 +1/4\alpha\beta^{-1}\gamma\vert\Lambda\vert \cosh(\sqrt{3\vert\Lambda\vert}r)
 \sinh^{ -2}(\sqrt{3\vert\Lambda\vert}r)
\\
\Psi_2 =  -1/8\alpha^{2}\beta^{ -2}\gamma^{2}\vert\Lambda\vert
  \sinh^{ -2}(\sqrt{3\vert\Lambda\vert}r)
 +1/3\vert\Lambda\vert \sinh^{-2}(\sqrt{3\vert\Lambda\vert}r)
\\
\phantom{\Psi_2 = }
 -1/6\delta\vert\Lambda\vert \cosh(\sqrt{3\vert\Lambda\vert}r)
  \sinh^{ -2}(\sqrt{3\vert\Lambda\vert}r) }
\end{equation}
\begin{equation}\eqalign {\label{CS2}
D\Psi_{01'} = D\Psi_{50'}\\
\phantom{D\Psi_{01'}}
 = \sqrt{6} \left( 1/16\alpha^{3}\beta^{-3}\gamma^{3}\vert\Lambda\vert^{3/2}
 \sinh^{-3}(\sqrt{3\vert\Lambda\vert}r)\right.
\\
\phantom{D\Psi_{01'} = }
-5/12\alpha\beta^{-1}\gamma\delta\vert\Lambda\vert^{3/2}
 \cosh(\sqrt{3\vert\Lambda\vert}r)\sinh^{ -3}
 (\sqrt{3\vert\Lambda\vert}r)
\\
\phantom{D\Psi_{01'} = }
 -5/24\alpha\beta^{ -1}\gamma\vert\Lambda\vert^{3/2}\sinh^{ -1}
 (\sqrt{3\vert\Lambda\vert}r)
\\
\phantom{D\Psi_{01'} = }
 \left. -2/3\alpha\beta^{ -1}\gamma\vert\Lambda\vert^{3/2}\sinh^{ -3}
 (\sqrt{3\vert\Lambda\vert}r) \right) }
\end{equation}
\begin{equation}\eqalign {\label{CS3}
D\Psi_{10'} = D\Psi_{41'}\\
\phantom{D\Psi_{01'}}
 = \sqrt{6} \left( -1/16\alpha^{3}\beta^{ -3}\gamma^{3}
 \vert\Lambda\vert^{3/2}\sinh^{-3}(\sqrt{3\vert\Lambda\vert}r)\right.
\\
\phantom{D\Psi_{01'} = }
 -1/12\alpha\beta^{-1}\gamma\delta\vert\Lambda\vert^{3/2}
 \cosh(\sqrt{3\vert\Lambda\vert}r)\sinh^{ -3}(\sqrt{3\vert\Lambda\vert}r)
\\
\phantom{D\Psi_{01'} = }
 -1/24\alpha\beta^{ -1}\gamma\vert\Lambda\vert^{3/2}\sinh^{ -1}
 (\sqrt{3\vert\Lambda\vert}r)
\\
\phantom{D\Psi_{01'} = }
 \left. +1/6\alpha\beta^{ -1}\gamma\vert\Lambda\vert^{3/2}\sinh^{ -3}
 (\sqrt{3\vert\Lambda\vert}r) \right)}
\end{equation}
\begin{equation}\eqalign {\label{CS4}
D\Psi_{21'} = D\Psi_{30'}\\
\phantom{D\Psi_{01'}}
 = \sqrt{6} \left( 1/8\alpha^{2}\beta^{-2}\gamma^{2}
  \vert\Lambda\vert^{3/2}\cosh(\sqrt{3\vert\Lambda\vert}r)\sinh^{ -3}
 (\sqrt{3\vert\Lambda\vert}r) \right.
\\
\phantom{D\Psi_{01'} = }
 +1/12\delta\vert\Lambda\vert^{3/2}\sinh^{ -1}(\sqrt{3\vert\Lambda\vert}r)
\\
\phantom{D\Psi_{01'} = }
 +1/6\delta\vert\Lambda\vert^{3/2}\sinh^{ -3}(\sqrt{3\vert\Lambda\vert}r)
\\
\phantom{D\Psi_{01'} = }
 \left. -1/3\vert\Lambda\vert^{3/2}\cosh(\sqrt{3\vert\Lambda\vert}r)\sinh^{
 -3}(\sqrt{3\vert\Lambda\vert}r) \right) }
\end{equation}
We see that the other essential parameter that appears is
$\alpha\gamma/\beta$ which is a linear combination of the $p_i$ and so
equivalent to specifiying $p_2$ as expected.
The results for the other possible cases described above are
similar in form. Thus, as expected, we find
that only two parameters are essential locally: all the other parameters
can be locally removed by coordinate transformations. 

We can also relate the Cartan scalars above to their limiting forms for
the Weyl vacuum class considered by da Silva et al.\ (1995b). In the
limit $\Lambda =0$ we can identify the other parameters in the Cartan scalars
with the $n$ of
the Weyl form by $\delta = 2(n^2-3)/(n^2+3)$, $\alpha\gamma/\beta =
8n/(n^2+3)$, and with these relations the Cartan scalars above have the da
Silva et al.\ values as their limits.

\section*{Acknowledgements}

We are grateful to the E.P.S.R.C. for a Visiting Fellowship grant for
N.O.S. to visit London, during which much of this work was completed,
and to Prof.\ G.F.R. Ellis of the University of Cape Town for
arranging support for a visit there by M.A.H.M. We are also much
indebted to Prof.\ W.B. Bonnor for many helpful comments and
criticisms on earlier drafts.


\end{document}